\documentclass[fleqn,usenatbib]{mnras}
\usepackage{newtxtext,newtxmath}
\usepackage{hyperref}
\usepackage{siunitx}
\usepackage{amsmath}
\usepackage{booktabs}
\usepackage{mathtools}
\usepackage{amsmath}
\usepackage{subfigure}
\usepackage{bm}
\usepackage{ulem}
\newcommand{\be}{\begin{equation}}
\newcommand{\ee}{\end{equation}}
\newcommand{\ba}{\begin{eqnarray}}
\newcommand{\ea}{\end{eqnarray}}
\newcommand{\bi}{\begin{itemize}}
\newcommand{\ei}{\end{itemize}}
\newcommand{\no}{\nonumber}

\usepackage{color}

\usepackage{ulem}

\let\oldAA\AA
\renewcommand{\AA}{\text{\normalfont\oldAA}}
\usepackage[T1]{fontenc}
\usepackage{xcolor}
\usepackage{bm}
\DeclareRobustCommand{\VAN}[3]{#2}
\let\VANthebibliography\thebibliography
\def\thebibliography{\DeclareRobustCommand{\VAN}[3]{##3}\VANthebibliography}
\usepackage{graphicx}
\usepackage{grffile}% Including figure files
\title{Numerical investigation of non-Gaussianities in the 
  phase and modulus of density Fourier modes}

\author[Qin et al.]{Qin Jian$^{1,2}$,
Pan Jun$^{4}$,
Yu Yu$^{1,2}$\thanks{E-mail:{yuyu22@sjtu.edu.cn}},
Zhang Pengjie$^{1,2,3}$\thanks{E-mail:{zhangpj@sjtu.edu.cn}}
\\
$^{1}$Department of Astronomy, School of Physics and Astronomy, Shanghai Jiao Tong University, Shanghai, 200240,China\\
$^{2}$Key Laboratory for Particle Astrophysics and Cosmology
(MOE)/Shanghai Key Laboratory for Particle Physics and Cosmology,
China\\
$^{3}$Tsung-Dao Lee Institute, Shanghai Jiao Tong University, Shanghai, 200240, China\\
$^{4}$National Astronomical Observatories, Chinese Academy of Sciences, Beijing 100101, China
}
\date{Accepted XXX. Received YYY; in original form ZZZ}

\begin{document}
\label{firstpage}
\pagerange{\pageref{firstpage}--\pageref{lastpage}}
\maketitle

\begin{abstract}
We numerically investigate non-Gaussianities in the late-time cosmological density field in Fourier space. We explore various
statistics, including the two-point and three-point probability
distribution function (PDF) of phase and modulus, and two \& three-point
correlation function of of phase and modulus. We detect significant
non-Gaussianity for certain configurations. We compare the simulation
results with the theoretical expansion series of
\citet{2007ApJS..170....1M}. We find that the $\mathcal{O}(V^{-1/2})$
order term alone
is sufficiently accurate to describe all the measured
non-Gaussianities in not only the PDFs, but also the correlations. We
also numerically find that the phase-modulus cross-correlation
contributes $\sim 50\%$ to the bispectrum, further verifying the
accuracy of the $\mathcal{O}(V^{-1/2})$ order prediction. This work
demonstrates that non-Gaussianity of the cosmic density field is
simpler in Fourier space, and may facilitate the data analysis in the
era of precision cosmology.  
\end{abstract}

%%%%%%%%%% BODY OF PAPER%%%%%
\section{Introduction}
The structure originates from the initial density
fluctuations which are thought to be (or at least nearly) Gaussian, under the inflation scenerios \citep{1981PhRvD..23..347G,1981PhLB...99...66S,1982PhLB..108..389L}.
The statistical properties of random Gaussian fields are
completely characterized by the two-point correlation function (or its Fourier counterpart, the power spectrum), which  
are very convenient and popular tool to characterize the large-scale
structure \citep{1980lssu.book.....P}. 
On the other hand, due to the gravitational evolution of the Universe, the late-time matter density fields are highly non-Gaussian \citep{2002PhR...367....1B,mellier1999probing,2001PhR...340..291B, 2008PhR...462...67M, Kilbinger2015Cosmology}.
As a result, cosmological information goes beyond the two-point statistics \citep{2005MNRAS.360L..82R,10.1111/j.1365-2966.2006.10710.x,2007MNRAS.375L..51N,2006MNRAS.370L..66N}.  
To effectively characterize and extract information form the late-time density fields, Cosmologists turn to non-Gaussian statistics such as the $n$-point correlation functions \citep[]{Bernardeau2002,Takada_2003, Semboloni_2011, Fu_2014}, the polyspectra \citep[e.g.,][]{Sefusatti_2006}, Minkowski functionals \citep{Mecke_1994, Hikage_2003, Shirasaki_2014, Kratochvil_2012,2020arXiv201200203M,2020arXiv201104954M}, counts of clusters, peaks, and voids \citep{Jain_2000, Marian_2009, Kratochvil_2010, LPH15, LPL15, 2018ApJ...862...60Q,Pisani_2019}, and statistics based on ideas with neural networks \citep[e.g.,][]{Gupta_2018, Ribli_2019a, Ribli_2019,2020MNRAS.499.5902C,2021arXiv210309247C}.
Applications of these non-Gaussian statistics to the large-scale structure of the Universe are quite popular \citep[e.g.,][]{2003PASJ...55..911H,2006ApJ...653...11H,2020PhRvD.101f3515L, 2010ApJ...719.1408F, 2011PhRvD..84d3529Y, 2016PhRvD..94d3533L, 2016PhRvL.117e1101L, 2016MNRAS.463.3653K, Shan2017KiDS, 2018MNRAS.478.5436G, 2018MNRAS.473.3190H, Martinet2018KiDS,2019JCAP...05..043C, 2019arXiv191004627M}.

When considering the n-point correlation function or the polyspectra, the number of configurations explodes with the number of n-points, which makes the measurement difficult. 
Methods to avoid the high complexity of the n-point functions are proposed. 
One is the local transformation, which can significantly reduce the non-Gaussianity in the field, and enhance the cosmological information in the two-point statistics \citep[e.g.,][]{1991MNRAS.248....1C,Neyrinck2009,Neyrinck2011,2016MNRAS.457.3652M,2011MNRAS.418..145J,2011PhRvD..84b3523Y,Simpson_2011, Carron_2013, Giblin_2018}. 
However,  recently \citet{2020ApJ...897..105Q} revealed that the copula is clearly non-Gaussian in the differential form. Since copula is invariant under local transformation, the above finding implies that local transformation can not reduce non-Gaussianity perfectly. On the other hand, in the Fourier space, the Line Correlation Function (LCF) relies on pure phase information to simplify the evaluation of non-Gaussianities \citep{2013ApJ...762..115O, 2015ApJ...804..132W,2017MNRAS.466.2496E}. 
The LCF is constructed from the phase correlations of the density field and has been applied, for instance, to constrain the growth rate of structure \citep{2020MNRAS.497.1765B}, to probe redshift-space distortions \citep{2019PhRvD..99j3530F,2015MNRAS.453..797E}, and to improve the cosmological constraints \citep{2018MNRAS.479.2743A,2017MNRAS.471.1581B}.

Theory of the Fourier mode probability distribution has been
constructed by \citet{2007ApJS..170....1M}. Based on the generating
function and the cumulant expansion theorem,
\citet{2007ApJS..170....1M} obtained the general expression of the
N-point Fourier mode distribution function, and the Edgeworth
expansion series. \citet{2004ApJ...600..553H} found that the leading
order non-Gaussian term is already sufficient to describe the 3-point
phase probability distribution function (PDF) in N-body
simulations. The LCF statistics is also based on the theory of   
\citet{2007ApJS..170....1M}. Recently, \cite{TBD1} investigated the growth of the Fourier mode moduli, providing an
alternative approach on the Fourier mode study. 

Given the importance of the \citet{2007ApJS..170....1M} theory in
undestanding the Fourier mode statistics, we carry out more
comprehensive numerical investigations. The Fourier mode statistics is
expanded into series of $\mathcal{O}(V^{-1/2,-1,\cdots})$. So the key
task is to quantify the relative difference of these
terms. We measure both the 2-point and 3-point phase PDFs. We also
derive and compare the N-point correlation functions of phases and
moduli. The major finding is that the expansion to
$\mathcal{O}(V^{-1/2})$ is sufficient for all the
investigated statistics. This agrees with the previous finding of
\citet{2004ApJ...600..553H} on the three-point phase PDF. 
Surveys such as BOSS, eBOSS, DESI and PFS have volumes of $\sim 10
(h^{-1}{\rm Gpc})^3$, much larger than the simulation volumes that we
investigate. Therefore a major conclusion is that  the $\mathcal{O}(V^{-1/2})$ order
non-Gaussianity expression of \citet{2007ApJS..170....1M} is
sufficiently accurate  for these surveys.  Another interesting finding
is that the phase-modulus cross-correlation is responsible for $\sim
50\%$ in the measured bispectrum.

This paper is organized as follows. In \S 2 we review the analytical formulas of the probability distribution functions of the Fourier modes, and we present the analytical predictions of the two and three point correlation functions of the phases and modulus derived from the PDFs. 
And we introduce our method to disrupt the phase-modulus correlation and investigate the influence on polyspectra.
In \S 3, we describe the simulation data, present the results
measured from simulations,  and compare to the analytical formulas.
We summarize the results in \S 4.

\section{ The DISTRIBUTION FUNCTION OF FOURIER MODES}\label{sec2}
Starting from the the cumulant expansion theorem of the Fourier
coefficients, \cite{2007ApJS..170....1M} derived the  general
expression of N-point probability distribution function of Fourier
modes $\delta(\bm k)=|\delta(\bm k)|e^{i\theta_{\bm k}}$.  We summary
the major results in  \cite{2007ApJS..170....1M}, and then derive the
major statistics that we will investigate in this paper. 

\subsection{Preliminaries of Fourier mode distribution}
The nonlinear evolution of the universes drives the n-point PDF

$\mathcal{P}_n$ of Fourier modes of $\delta({\bf k}_j)$
($j=1,\cdots,n$) to deviate from the Gaussian PDF
$\mathcal{P}_{G,n}$.  Following \citet{2007ApJS..170....1M}, the
arguments of PDF are taken as $A_1,\theta_1,\cdots, A_n,\theta_n$. Here $A_i\equiv A_{\boldsymbol{k}_i}$ is  the
normalized Fourier modulus ($A_{\bm{k}} \equiv |\delta(\bm k)|/\sqrt{P(\bm{k})}$) and 
$\theta_i\equiv \theta_{\boldsymbol{k}_i}$ is  the phase. For brevity,
we often ignore the arguments in $\mathcal{P}_{n}$ and
$\mathcal{P}_{G,n}$.  The Gaussian
PDF 
\begin{equation}
\label{PG}
\mathcal{P}_{G,n}=\prod_{i=1}^n P_i\ ,\ P_i=A_ie^{-A_i^2} \pi^{-1}\ .
\end{equation}
As we have learned from the central limit theorem, non-Gaussian corrections to 
$\mathcal{P}_n$ decrease with the
number of independent Fourier modes. Therefore they have explicit
dependence on the volume $V$ in measuring the
Fourier mode.  
\citet{2007ApJS..170....1M} derived the full expression (its Eq. 48) in
series of $V^{-1/2,-1,\cdots}$,
\ba
\frac{\mathcal{P}_n}{\mathcal{P}_{G,n}}-1=\sum_{i=1}^{\infty} C_{-i/2}\ .
\ea
$C_{-i/n}$ is the non-Gaussian correction of $V^{-i/2}$ dependence. So
in general case, the leading order non-Gaussian correction is
\ba
\label{eq:1st-order-appr}
C_{-1/2}&=&V^{-1/2} \\
&\times&\sum_{{\boldsymbol{k_1}},{\boldsymbol{k_2}},{\boldsymbol{k_3}}}^{\rm
  uhs}A_1A_2A_3\cos(\theta_1+\theta_2-\theta_3)
p^{(3)}({\boldsymbol{k_1}},{\boldsymbol{k_2}},-{\boldsymbol{k_3}})\ . \no
\ea
The symbol ``uhs'' indicates the ``upper half sphere'': $k_z\ge 0$ of the $\bm{k}$-space.
The bispectrum is defined through
\begin{equation}
\begin{aligned}
   \left\langle \delta(\bm k_1) \delta(\bm k_2)\delta(\bm k_3) \right\rangle &=
   V^{-1/2}\delta^{\rm K}_{\bm k_1 + \bm k_2+ \bm k_3} P^{(3)}(\bm{k}_1,\bm{k}_2,\bm{k}_{3})\ ,\\
  p^{(3)}(\bm{k}_1,\bm{k}_2,\bm{k}_3) &\equiv
  \frac{\delta^{\rm K}_{\bm{k}_1 + \bm{k}_2 + \bm{k}_3}
     P^{(3)}(\bm{k}_1,\bm{k}_2,\bm{k}_{3})}
    {\sqrt{P(\bm{k}_1)P(\bm{k}_2)P(\bm{k}_{3})}}\ .
\end{aligned}
\label{eq:2-22}
\end{equation}
Notice that the normalized bispectrum $p^{(3)}(\bm{k}_1,\bm{k}_2,\bm{k}_3)$ does not depend on $V$, so the non-Gaussian correction to $\mathcal{P}$ decreases with increasing volume, at the rate of $\propto V^{-1/2}$.

The next-to-leading order non-Gaussian term $C_{-1}$ has a $V^{-1}$
dependence. So it is sub-dominant to $C_{-1/2}$ when $C_{-1/2}\neq
0$. However, for some configurations $C_{-1/2}=0$
(e.g. $\mathcal{P}_3$ with ${\bf k}_1+{\bf k}_2+{\bf k}_3\neq 0$) and
$C_{-1}$ will be the dominant contribution of
non-Gaussianity. $C_{-1}$ depends on both the bispectrum and
trispectrum and the full expression is given by Eq. 58 of
\citet{2007ApJS..170....1M}. 

\subsection{Derived statistics}
The above results allow us to derive various statistics, such as the
phase distribution and phase correlation function, modulus
distribution and modulus correlation function, and also the
phase-modulus cross-correlation.  
\subsubsection{Phase distribution and correlation function}
Using Eq. \ref{eq:1st-order-appr}, we can obtain the phase PDF $\mathcal{P}(\theta_1,\cdots,\theta_n)$, 
\begin{equation}
\begin{aligned}
\label{eq:1st-order-appr-ph}
\frac{\mathcal{P}(\theta_1,\cdots,\theta_n)}{\mathcal{P}_{G}(\theta_1,\cdots,\theta_n)}=1&+\frac{\pi^{3/2}}{4\sqrt{V}}\sum_{{\boldsymbol{k_1}},{\boldsymbol{k_2}},{\boldsymbol{k_3}}}^{\rm uhs}\cos\left(\theta_1+\theta_2-\theta_3\right)\\
	& \times p^{(3)}({\boldsymbol{k_1}},{\boldsymbol{k_2}},-{\boldsymbol{k_3}})+{\mathcal O}(V^{-1})\ ,
\end{aligned}
\end{equation}
Here $\mathcal{P}_{G}(\theta_1,\cdots,\theta_n)=(2\pi)^{-n}$. 
Therefore in general the phase distribution show non-Gaussianity at
the order of $V^{-1/2}$.  We can then derive the n-point phase
correlation functions. For example, the 2-point phase correlation
\begin{equation}
\begin{aligned}
\label{eq:2ptphph}
\left\langle\theta_{\bm k}\theta_{2\bm k}\right\rangle
&=\left \langle\theta_{\bm k}\theta_{2\bm k}\right\rangle_{\rm G}
+\frac{\sqrt\pi}{4\sqrt{V}}p^{(3)}\left(\bm{k},\bm{k},-2\bm{k}\right)\\
&=\pi^2+\frac{\sqrt\pi}{4\sqrt{V}}p^{(3)}\left(\bm{k},\bm{k},-2\bm{k}\right)
+ \mathcal O(V^{-1})\ . 
\end{aligned}
\end{equation}
The phase correlation function of 3 Fourier modes 
\begin{eqnarray}
\label{eq:3ptphph}
\left\langle\theta_1^{m_1}\theta_2^{m_2}\theta_{12}^{m_3}\right\rangle
&=\left\langle\theta_{1}^{m_1}\theta_2^{m_2}\theta_{12}^{m_3}\right\rangle_{\rm G}
+\left\langle\theta_1^{m_1}\theta_2^{m_2}\theta_{12}^{m_3}\right\rangle_{\rm NG} \\
&=\frac{(2\pi)^{m_1+m_2+m_3}}{(m_1+1)(m_2+1)(m_3+1)}
+\left\langle\theta_1^{m_1}\theta_2^{m_2}\theta_{12}^{m_3}\right\rangle_{\rm NG}
  \ .\nonumber
\end{eqnarray}
Here $\theta_{12}\equiv \theta_{{\bf k}_1+{\bf k}_2}$. The
non-Gaussian contribution
$\langle\theta_1^{m_1}\theta_2^{m_2}\theta_{12}^{m_3}\rangle_{\rm NG}$
is generally nonzero. For instance, 
\begin{eqnarray}
\label{eq:xi3ptphph1}
\left\langle\theta_1\theta_2\theta_{12}^{2}\right\rangle
=\frac{4\pi^{4}}{3}
+\frac{\pi^{3/2}}{2\sqrt{V}}
p^{(3)}({\bm k_1},{\bm k_2},{-\bm k_1-\bm k_2})\\ \nonumber
+ \mathcal O(V^{-1})\ .
\end{eqnarray}
The correlation function $\langle e^{i(\theta_1+\theta_2-\theta_{12})}\rangle$ combines these polynomials and has 
\begin{eqnarray}
\label{eq:3ptphph2}
\left\langle e^{i(\theta_1+\theta_2-\theta_{12})}\right\rangle=\frac{\pi^{3/2}}{8\sqrt{V}}
p^{(3)}({\bm k_1},{\bm k_2},{-\bm k_1-\bm k_2})\\ \nonumber
+ \mathcal O(V^{-1})\ .
\end{eqnarray}

\subsubsection{Modulus distribution and correlation function}
In contrast, the modulus distribution is much more Gaussian,  since from Eq. \ref{eq:1st-order-appr} we have
\begin{equation}
\begin{aligned}
\label{eq:jpdfAk}
\frac{\mathcal{P}(A_1,\cdots,A_n)}{\mathcal{P}_{\rm G}(A_1,\cdots,A_n)}=1+{\mathcal O}(V^{-1})\ .
\end{aligned}
\end{equation}
So for mild non-Gaussian field, we expect the modulus distribution can be well approximated by the N-point independent Rayleigh distribution,
\begin{equation}
 \mathcal{P}_G(A_1,\cdots,A_n)= \prod_i 2 A_i e^{-A_i^2}\ .
\label{eq:jpdfAkG}
\end{equation}
These barely correlated modulus combined with random phases (we call
it the randomization of phases) yield Gaussian fields. 

We can also derive the N-point correlation functions of the Fourier coefficients from the distribution, such as the 2-point moduli correlation
\begin{equation}
\begin{aligned}
\label{eq:xi2ptmodmod1}
\left\langle A_{\bm k}^mA_{2\bm k}^n\right\rangle
=\left \langle A^m_{\bm k}A^n_{2\bm k}\right\rangle_{\rm G}
+\mathcal O(V^{-1})=I(m)I(n)+\mathcal O(V^{-1})\ .
\end{aligned}
\end{equation}
Here
\begin{equation}
	 I(n) \equiv \int_0^\infty A^n \cdot 2A e^{-A^2} dA =
  \Gamma\left(1 + \frac{n}{2}\right)\nonumber
\end{equation}

At the oder of $V^{-1}$ and above, there are non-vanishing
non-Gaussian corrections to $\mathcal{P}(A)$, and the modulus
correlation functions. These corrections are evaluated in \S
\ref{sec:PDF}. 
\subsubsection{Phase-modulus cross correlation}
Nevertheless, the nearly Gaussian modulus distribution does not
indicate that the modulus contains no non-Gaussian information. The
point is that the moduli are correlated with phases.
It is interesting to ask how much this modulus-phase correlation
contribute to the non-Gaussianity. This can be quantified by the
following operation. We generate a new field with phase correlations
retained, but modulus-phase correlations removed by replacing the
original moduli with random variables generated by the Rayleigh distribution. Then from equation \eqref{eq:1st-order-appr-ph} the joint distribution of the Fourier modes of the new field has
\begin{equation}
\label{PG3}
\frac{\mathcal{P}^\prime_n}{\mathcal{P}_{G,n}}=1+\sum_{{\boldsymbol{k_1}},{\boldsymbol{k_2}},{\boldsymbol{k_3}}}^{\rm uhs}\frac{\pi^{3/2}}{4\sqrt{V}}\cos(\theta_1+\theta_2-\theta_3)p^{(3)}({\boldsymbol{k_1}},{\boldsymbol{k_2}},-{\boldsymbol{k_3}})
\end{equation}
Integration over $\mathcal{P}^\prime$ yields the bispectrum of the new field
\begin{equation}
\begin{aligned}
\label{eq:newbs}
p^{\prime(3)}(\bm k_1 ,\bm k_2 ,-\bm k_3)=\left(\frac{\pi}{4}\right)^3\times
p^{(3)}(\bm k_1 ,\bm k_2 ,-\bm k_3)\ .
\end{aligned}
\end{equation}
We can see that there is a factor of $(\pi/4)^3\simeq 0.484$ decrease of the bispectrum if we decorrelate the phase-modulus. This empirically reflects the contribution to non-Gaussianity of the phase-modulus correlation.

To sum up, the modulus distribution of Fourier modes are nearly
Gaussian and  largely independent to each other.   The leading order
non-Gaussian correction is of the order $V^{-1}$. In contrast, the
phase distribution is significantly more non-Gaussian. The leading
order non-Gaussian correction, for general cases, is of the order
$V^{-1/2}$ and is proportional to the bispectrum. In this work, we will numerically
  test these results for the cosmological density fields, including:
\begin{itemize}
  \item The PDF of the Fourier modulus. 
  \item The correlation functions of Fourier moduli. Direct comparison of the PDF may not be clear in terms of accuracy, especially for two or more point cases. The comparison of correlation functions can show differences more clearly in these cases.   
  \item The PDF of the phase closure (e.g., $\theta_1+\theta_2-\theta_{12}$). To first order, the joint PDFs of phases are in the form of phase closures, which we compare to the distribution measured from simulated density fields.
  \item The phase correlation functions, which are  statistics complementary to the phase closure distributions.   
\end{itemize}

\begin{figure} 
%\hspace{-2cm}
\centering
\includegraphics[width=9.5cm]{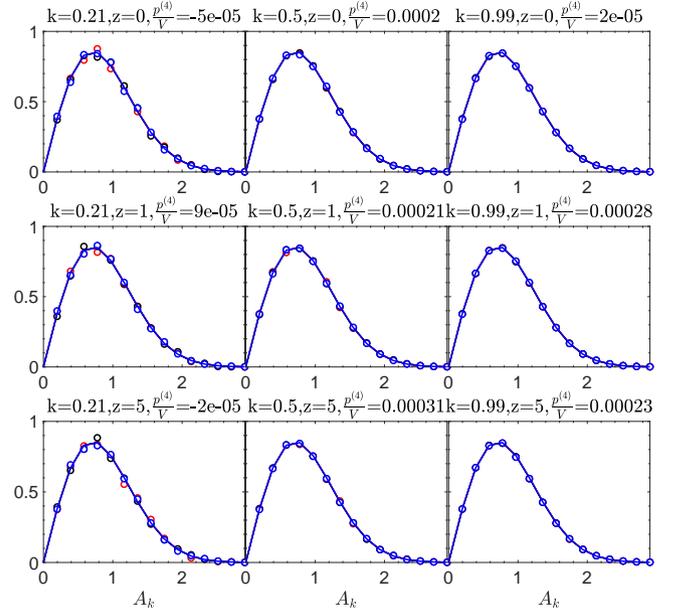}
%\vspace{-1.5cm}
\caption{The one-point distribution function of normalized Fourier modulus of the density fields at three redshifts (from top to bottom panels). 
Three cases of wave number $k \sim 0.2{\rm Mpc}/h$, $0.5{\rm Mpc}/h$ and $1.0{\rm Mpc}/h$  (averaged with bin width $0.1{\rm Mpc}/h$) are shown from left to right panels, respectively. The solid lines show the Rayleigh distribution and the dotted lines show the non-Gaussian prediction Eq. \eqref{eq:1ptpdfAkNG}. The three colors represent the results from the three realizations of the simulation. The non-Gaussian prediction is not distinguishable from the Rayleigh distribution.  
\label{fig:1ptmodule}}
\end{figure}

\section{Results}
\label{sec:PDF}
\subsection{Simulation data}
\label{sec:method}
The simulations we use for the test were run with $3072^3$ particles in a box with volume $V=(600\  {\rm Mpc}/h)^3$, and a flat cosmology specified by $\Omega_m = 0.268$, $\Omega_\Lambda = 0.732$,
$H_0 = 71\ \rm{km}\ s^{-1}{Mpc}^{-1}$, $\sigma_8 = 0.83$, $n_s =
0.968$.  The details of the simulations and the CosmicGrowth
simulation series are described in \cite{2007ApJ...657..664J} \& \cite{2019SCPMA..6219511J}.
The density fields are sampled at redshifts $z=0,1,5$ with the grid
size $1 {\rm Mpc}/h$. The mean number of particles per grid is
$134.2$, so we can safely neglect the effect of shot noise.  The
simulations have three different realizations, allowing us to better
handle the cosmic variance. 

\begin{figure} 
%\hspace{-2cm}
\includegraphics[width=9cm]{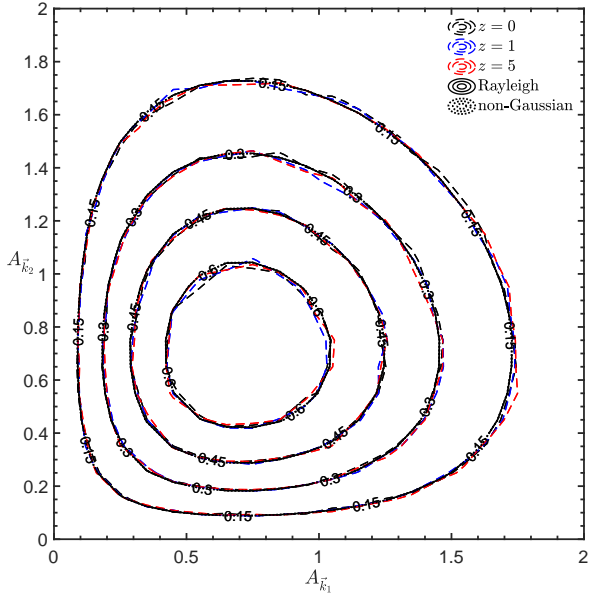}
%\vspace{-1.5cm}
\caption{Contour plots of the 2-point moduli PDF at redshifts $z=0$ (black dashed lines), $z=1$ (blue dashed lines) and $z=5$ (red dashed lines). The distribution are measured for $\bm k_2=2\bm k_1$ and $|\bm k_1|$ in range of $[0.9,1.1]{\rm Mpc}/h$.
    The black solid lines show the Rayleigh distribution and dotted lines show the non-Gaussian prediction.  
\label{fig:jpdfmodule}}
\end{figure}

\subsection{Fourier modulus}
Fig. \ref{fig:1ptmodule} shows the one-point PDF of the normalized
Fourier modulus, for $z=0,1,5$ and 
three $k$ bins at $[0.1,0.3]$, $[0.4,0.6]$ and $[0.9,1.0]$ (in units
of $h/$Mpc).  It is very close to Gaussian. The results indicates the
central limit theory, i.e., the simulation box is large such that the
Fourier transform sums over a large number of independent modes. This
behavior has also been predicted by \cite{1995PhRvD..51.6714F}. The leading order
non-Gaussian correction is
\be
  \frac{\mathcal{P}(A)}{\mathcal{P}_G(A)}-1 = \frac{1}{V}
    \left(
      \frac14 A^4 - A^2 + \frac12
    \right)
     p^{(4)}\left(\bm{k},\bm{k},-\bm{k},-\bm{k}\right)\ . 
\label{eq:1ptpdfAkNG}
\ee
We use the trispectrum measured from the same simulations to calculate
this correction and plot it in Fig. \ref{fig:1ptmodule}. We find that it is totally
negligible. 

Fig. \ref{fig:jpdfmodule} shows the 2-point moduli PDF
$\mathcal{P}(A_1,A_2)$ of ${\bf k}_1$
and ${\bf k}_2=2{\bf k}_1$. Again the PDF is very close to
Gaussian. The leading order non-Gaussian correction,  $\mathcal{P}_2/\mathcal{P}_{G,2}-1$ is 
\ba
\frac{1}{2V}&\times&
  \left\{\left[
 -2A_1^2 A_2^2 -
      \frac12 A_1^4 +
      2 A_1^2 + A_2^2 - 1
  \right] \right.\no \\
&&\left. 
  \left[
    p^{(3)}\left(\bm{k}_1,\bm{k}_1,-\bm{k}_2\right)
  \right]^2
    + {\rm sym.}\left(\bm{k}_1 \leftrightarrow \bm{k}_2\right) \right\}
\nonumber\\
+ \frac{1}{V}&\times&
    \left(
      A_1^2 A_2^2 -
      A_1^2 - A_2^2 + 1
    \right)
    p^{(4)}\left(\bm{k}_1,\bm{k}_2,-\bm{k}_1,-\bm{k}_2\right). \nonumber \\ 
\label{eq:2ptjpdfAk}
\ea
Figure \ref{fig:jpdfmodule} also shows the prediction including the
above corrections. Again, their impacts are negligible and the 2-point
PDF of moduli is well described by the Gaussian Fourier mode
distribution.

The above PDF may not be most suitable to demonstrate the non-Gaussian
corrections. So we compare the moduli correlation functions between
two Fourier modes, which is
a compressed version of the two-point moduli PDF. We only investigate
the cases of ${\bm k}_2=2{\bm k}_1$ and $z=0$. But we investigate different orders in $A$
($\langle A_{\bm k}^mA_{2\bm k}^n\rangle$), with
$(m,n)=(1,1),(1,2),(2,2)$. For all cases, the simulation results agree
well with the Gaussian prediction (Eq. \ref{eq:xi2ptmodmod1}).  

The leading order non-Gaussian term,  $\mathcal{O}(V^{-1})$, is
\begin{eqnarray}
  \frac{1}{2V}&\times&
  \left[
 -2 I(m+2)I(n+2) -
      \frac12 I(m+4) \right. \no \\
&&+\left.
      2 I(m+2) + I(n+2) - 1\right]
  \left[
    p^{(3)}\left(\bm{k},\bm{k},-2\bm{k}\right)
  \right]^2
\nonumber\\
  + \frac{1}{V}&&
    \left[
      I(m+2) I(n+2) -
      I(m+2) - I(n+2) + 1
    \right]\no\\
&&
    \times p^{(4)}\left(\bm{k},2\bm{k},-\bm{k},-2\bm{k}\right). 
\label{eq:xi2ptmodmod2nd}
\end{eqnarray}

\begin{figure} 
%\hspace{-2cm}
\includegraphics[width=9.15cm]{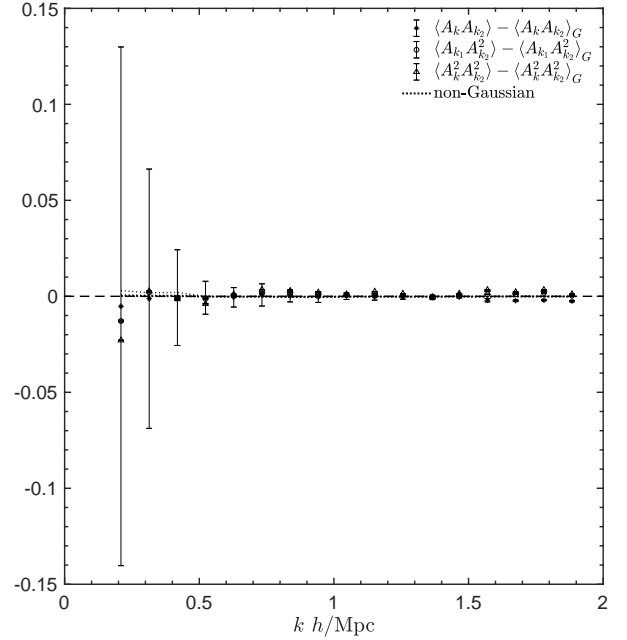}
%\vspace{-1.5cm}
\caption{Moduli correlation functions $\langle A_{\bm k}^mA_{2\bm k}^n\rangle$ as a function of $k=|\bm k|$ at redshift $z=0$. 
The results have substracted the zero order (Gaussian) term.
The dotted lines represent the non-Gaussian prediction to the $\mathcal{O}(V^{-1})$ order. 
The non-Gaussian correction is too tiny to be visible for most cases.  The errorbars are measured from downsample method.
\label{fig:ximodmod}}
\end{figure}
Figure \ref{fig:ximodmod} shows that the non-Gaussian correction is
tiny. To quantify its impact, we randomly divide each $k$ bin into 8
subsamples to estimate the errorbars shown in Fig
\ref{fig:ximodmod}. Within the simulation statistical errors, the
non-Gaussian correction is negligible in the moduli correlation
function. So the major conclusion is that non-Gaussian corrections to the moduli PDF and
correlation functions are all negligible. 

\subsection{Fourier phases}
\begin{figure} 
%\hspace{-1.65cm}
\centering
\includegraphics[width=9cm]{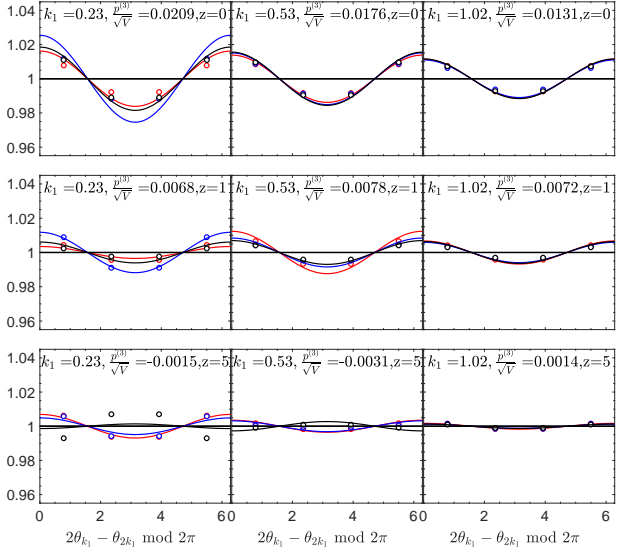}
%\vspace{-1.5cm}
\caption{The reduced two-point distribution (data points) of the Fourier phases of the density fields for $\bm k_2=2\bm k_1$ at three redshifts (from top to bottom panels). Cases for $|\bm k_1|=0.23{\rm Mpc}/h$, $0.53{\rm Mpc}/h$ and $1.02{\rm Mpc}/h$ are shown from left to right panels, respectively. $|\bm k_1|$ is averaged from bins of width $0.1{\rm Mpc}/h$. The three colors represent the three realizations of the simulation.
The solid lines represent analytical prediction to the $\mathcal{O}(V^{-1/2})$ order and dotted lines to the $\mathcal{O}(V^{-1})$ order. The dotted lines are not distinguishable from the solid lines.
\label{fig:2ptphase1}}
\end{figure}

For the phase PDF, the one-point PDF is always flat
\citep{2007ApJS..170....1M}. So we start with the 2-point PDF to
examine the non-Gaussian corrections. 
\subsubsection{Two-point} 
Fig. \ref{fig:2ptphase1} shows the phase PDF of $\bm k_1=2\bm
k_2$. Firstly, it shows statistically significant deviation from the
Gaussian case (flat PDF). The deviation is well described by the
leading order non-Gaussian correction ($\mathcal{O}(V^{-1/2})$),
\begin{equation}
\label{eq:2ptphjpdf}
 \mathcal{O}(V^{-1/2})= \frac{\sqrt{\pi}}{2\sqrt{V}}
  \cos(2\theta_{\bm{k}} - \theta_{2\bm{k}})
  p^{(3)}\left(\bm{k},\bm{k},-2\bm{k}\right) \ .
\end{equation}
We also calculate the second order
correction ($\mathcal{O}(V^{-1}$)
\begin{align}
  O(V^{-1}) =
  \frac{1}{2V}
  \cos\left[2(2\theta_{\bm{k}} - \theta_{2\bm{k}})\right]
  \left[
    p^{(3)}\left(\bm{k},\bm{k},-2\bm{k}\right)
  \right]^2\ .
\label{eq:2ptphpdf2nd}
\end{align}
Fig. \ref{fig:2ptphase1} shows that the $\mathcal{O}(V^{-1})$ is
negligible comparing to the leading $\mathcal{O}(V^{-1/2})$ term. 

\begin{figure} 
\centering
\includegraphics[width=9.5cm]{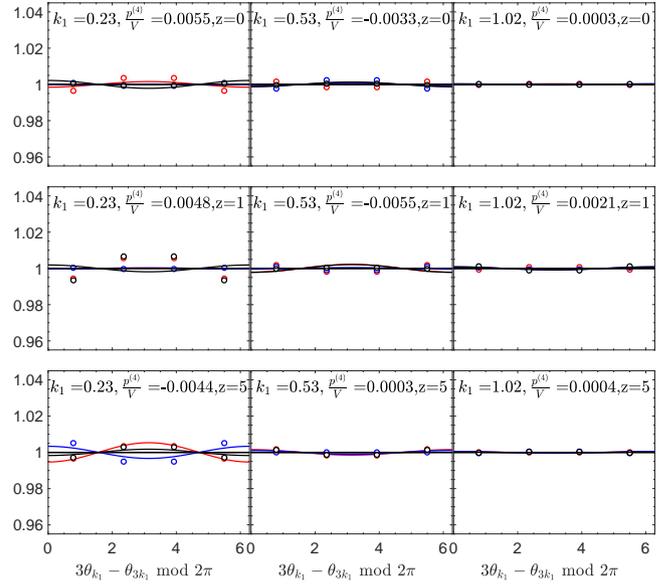}
\caption{Same as Fig \ref{fig:2ptphase1} but for configuration $\bm k_2=3\bm k_1$ and the solid lines represent analytical prediction to the $\mathcal{O}(V^{-1})$ order. To the $\mathcal{O}(V^{-1/2})$ order a uniform distribution is expected.
\label{fig:2ptphase2}}
\end{figure}

The situation is different for other configurations, such as $\bm
k_1=3\bm k_2$. In such case, the $\mathcal{O}(V^{-1/2})$ vanishes, and
the leading order non-Gaussian correction is  
\ba
  O(V^{-1}) =
   \frac{\pi}{8V}
    \cos(3\theta_{\bm{k}} - \theta_{3\bm{k}})
    p^{(4)}\left(\bm{k},\bm{k},\bm{k},-3\bm{k}\right)
\label{eq:2ptphph2nd}
\ea
Fig. \ref{fig:2ptphase2} shows that in such case, the non-Gaussian
correction in both simulation and theory are small, and can be
neglected within the simulation statistical uncertainties.  For other
configurations (${\bm k}_2\neq 2{\bm k}_1$ and ${\bm k}_2\neq
3{\bm k}_1$), non-Gaussian corrections only show in the order of
$\mathcal{O}(V^{-3/2})$ or above. Therefore for these configurations,
phases can be treated as uncorrelated. 
%%%%%%%%%%%%%%%5
%%%%%%%%%%%%%%%

\begin{figure} 
%\hspace{-1.62cm}
\centering
\includegraphics[width=9.25cm]{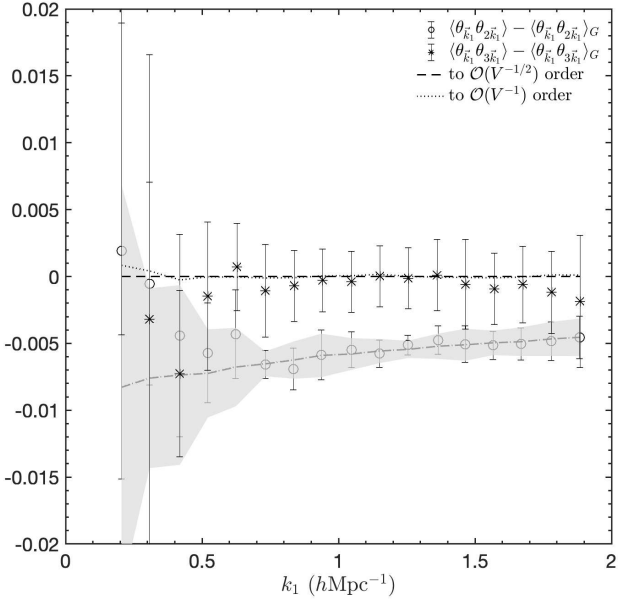}
%\vspace{-1.5cm}
\caption{The two-point phase correlation functions for configaration $\bm k_2=2\bm k_1$ (round dots) and $\bm k_2=3\bm k_1$ (star dots) at redshift $z=0$. 
The results have substracted the zero order (Gaussian) terms.
The dashed lines represent analytical prediction to the $\mathcal{O}(V^{-1/2})$ order and dotted lines to the $\mathcal{O}(V^{-1})$ order. The $\mathcal{O}(V^{-1})$ order correction is tiny for both configuarations. 
The errorbars are measured from downsample method.
The grey shadow regions in the $\bm k_2=2\bm k_1$ case indicate the fluctuation of the analytical predictions among the downsamples.
\label{fig:xiphph2pt}}
\end{figure}

Then we compare the correlation functions. For ${\bf k}_2=2{\bf k}_1$, 
\begin{equation}
\begin{aligned}
\label{eq:2ptxiphph2k}
\langle\theta_{\bm k}\theta_{2\bm k}\rangle-\pi^2
&=\frac{\sqrt\pi}{4\sqrt{V}}p^{(3)}\left(\bm{k},\bm{k},-2\bm{k}\right)
+ \mathcal O(V^{-1})
\end{aligned}
\end{equation}
with,
\begin{align}
  \mathcal O(V^{-1}) =
  \frac{1}{16V}
  \left[
    p^{(3)}\left(\bm{k},\bm{k},-2\bm{k}\right)
  \right]^2\ .
\label{eq:2ptxiphph2nd}
\end{align}
But for ${\bf k}_2=3{\bf k}_1$, the leading order correction is
$\mathcal{O}(V^{-1})$, 
\ba
\label{eq:2ptxiphph3k}
\langle\theta_{\bm k}\theta_{3\bm k}\rangle-\pi^2
=
  \frac{\pi}{24V}
    p^{(4)}\left(\bm{k},\bm{k},\bm{k},-3\bm{k}\right)+\cdots
\ea
For this reason, we do not detect the impact of non-Gaussianity in the
case of  ${\bf k}_2=3{\bf k}_1$ (Fig. \ref{fig:xiphph2pt}). In contrast, the detection of non-Gaussianity in the configaration
$\bm k_2=2\bm k_1$ is more significant at most $k$ ranges. The
deviation measured from the simulation (Fig. \ref{fig:xiphph2pt}) agrees excellently with the
theoretical prediction (Eq. \ref{eq:2ptxiphph2k},
$\mathcal{O}(V^{-1/2})$). 
 The
possible exception is at small $k$. But given the relatively less
Fourier modes, the statistical fluctuations are larger and the
deviation from Gaussian is statistically insignificant.

\subsubsection{Three-point}
The $\mathcal{O}(V^{-1/2})$ order correction to the three-point phase
PDF only exists for the configuration $\bm k_1+\bm k_2=\bm k_3$. For
it, the fractional corrections are  
\be
\label{eq:3ptphjpdf1}
\mathcal{O}(V^{-1/2})=\frac{\pi^{3/2}}{4\sqrt{V}}
  \cos(\theta_1+\theta_2 - \theta_3)
  p^{(3)}\left(\bm k_1,\bm k_2,-\bm k_3\right)\ ,
\ee
and the next-to-leading order correction is 
\be
  \mathcal O(V^{-1}) =
  \frac{1}{V}
  \cos\left[2(\theta_1+\theta_2 - \theta_3)\right]
  \left[
    p^{(3)}\left(\bm k_1,\bm k_2,-\bm k_3\right)
  \right]^2\ .
\label{eq:3ptphjpdf1_2nd}
\ee
For $2\bm k_1\pm\bm k_2=\bm k_3$, the leading order (fractional) correction is 
\begin{align}
  \mathcal O(V^{-1}) =
   \frac{\pi}{4V}
    \cos(2\theta_1 \pm \theta_2 - \theta_3)
    p^{(4)}\left(\bm k_1,\bm k_1,\pm\bm k_2,-\bm k_3\right)\ .
\label{eq:3ptphjpdf2_2nd}
\end{align}
For other configurations,  non-Gaussian corrections are of the order
$\mathcal{O}(V^{-3/2})$ or higher, and are negligible.   
\begin{figure} 
%\hspace{-1.65cm}
\centering
\includegraphics[width=9cm]{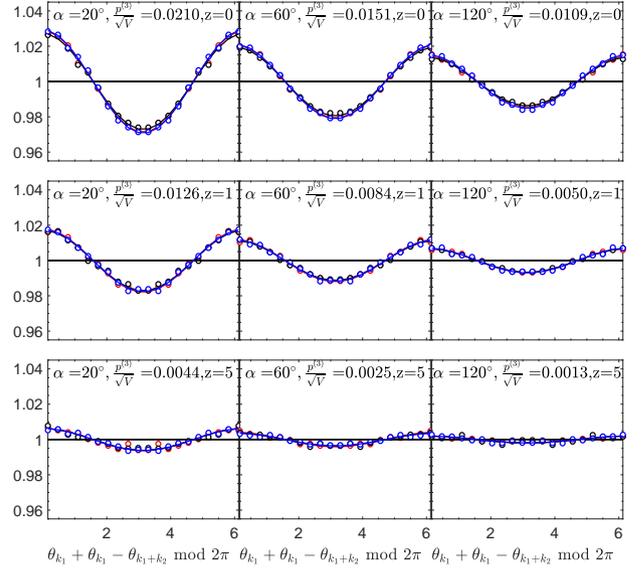}
%\vspace{-1.5cm}
\caption{The reduced three-point phase distribution (data points) of the Fourier phases of the density fields for configuration $
\bm k_1 +\bm k_2=\bm k_3$ at the three redshifts (from top to bottom panels). Three cases of open angles $\alpha$ are shown from left to right panels. The three colors represent the three realizations of the simulation.
The solid lines show the prediction to the $\mathcal{O}(V^{-1/2})$ order and dotted lines to the $\mathcal{O}(V^{-1})$ order.
\label{fig:3ptphase1}}
\end{figure}

\begin{figure} 
%\vspace{-2cm}
\centering
\includegraphics[width=9.5cm]{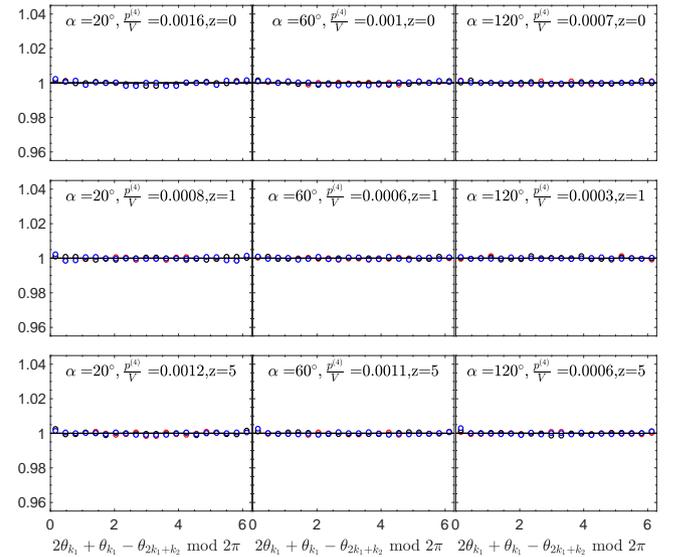}
%\vspace{-1.5cm}
\caption{Same as Fig \ref{fig:3ptphase1} but for configuration $2\bm k_1 +\bm k_2=\bm k_3$ and the solid lines represent analytical prediction to the $\mathcal O(V^{-1})$ order. 
\label{fig:3ptphase2}}
\end{figure}

Fig. \ref{fig:3ptphase1} \& \ref{fig:3ptphase2} shows the reduced three-point distribution of the Fourier phases for configuration $\bm k_1 + \bm k_2=\bm k_3$ and $\bm k_1 + 2\bm k_2=\bm k_3$ respectively. 
For both cases, the distribution are measured from the configuration
that $|\bm k_1|$ and $|\bm k_2|$ lie in range $[0.5,0.6]\ h^{-1}{\rm
  Mpc}/h$ and open angle $\alpha$ between $\bm k_1,\bm k_2$ lie
in range $[19\degr,21\degr],[59\degr,61\degr]$ and
$[119\degr,121\degr]$. For $\bm k_1 + \bm k_2=\bm k_3$, the
non-Gaussian correction is significant (Fig. \ref{fig:3ptphase1}). Due
to the large sample size of $k$ modes, the detection of
non-Gaussianity is improved comparing to the case of two-point
distribution. The
measurement agrees well with the
leading order correction in theory ($\mathcal{O}(V^{-1/2})$). The
next-to-leading order correction ($\mathcal{O}(V^{-1})$ is
negligible. 

For $2\bm k_1+\bm k_2=\bm k_3$, we detect no significant deviation from
the Gaussian distribution. This is also expected in the theory (Eq. \ref{eq:3ptphjpdf1_2nd}).
For $2\bm k_1+\bm k_2=\bm k_3$ the prediction is determined by a factor of $\frac{p^{(4)}}{V}$ (Eq. \ref{eq:3ptphjpdf2_2nd}) which are  marked out in each panels of Fig \ref{fig:3ptphase2}.

We then expect the analytical formulas to describe the  3-point
phase correlation functions, to a better accuracy than the 2-point cases.
Taking configuration $
\bm k_1 +\bm k_2=\bm k_3$ as an example, from Eq. \ref{eq:3ptphjpdf1} we have
\be
\label{eq:xi3ptphph1}
\langle\theta_1\theta_2\theta_{12}^{2}\rangle-\frac{4\pi^{4}}{3}=
\frac{\pi^{3/2}}{2\sqrt{V}}
p^{(3)}({\bf k}_1,{\bf k}_2, -{\bf k}_1-{\bf k}_2)
+ \mathcal O(V^{-1})\ ,
\ee
\begin{align}
  \mathcal O(V^{-1}) =
  \frac{1}{8V}
  \left[
    p^{(3)}\left(\bm k_1,\bm k_2,-\bm k_1-\bm k_2\right)
  \right]^2\ .
\label{eq:2ptxiphph2nd}
\end{align}

Another correlation is 
\begin{equation}
\begin{aligned}
\label{eq:xi3ptphph2}
\langle e^{i(\theta_1+\theta_2-\theta_{12})}\rangle
&=\frac{\pi^{3/2}}{8\sqrt{V}}
p^{(3)}({\bf k}_1,{\bf k}_2, -{\bf k}_1-{\bf k}_2)
+ \mathcal O(V^{-1})\ .
\end{aligned}
\end{equation}
But now the $\mathcal O(V^{-1})$ term is exactly zero. 

\begin{figure} 
%\hspace{-1.62cm}
\centering
\includegraphics[width=9cm]{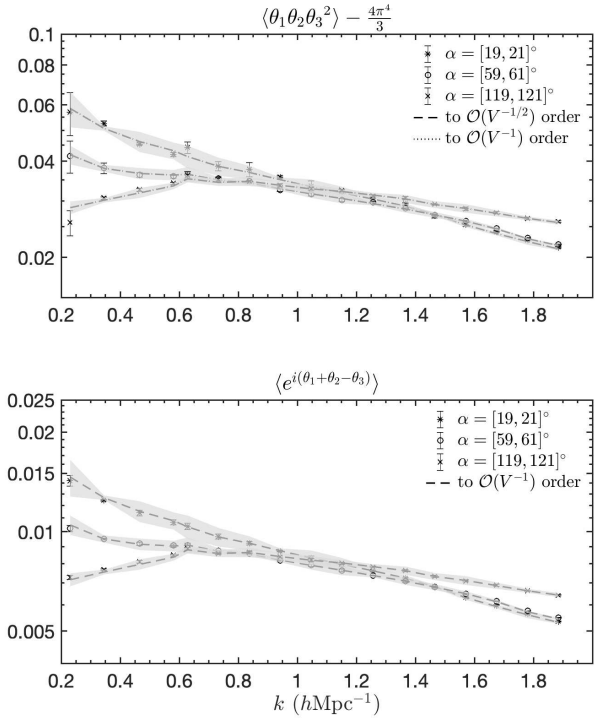}
%\vspace{-1.5cm}
\caption{The three-point phase correlation functions for configuration $
\bm k_1 +\bm k_2=\bm k_3$, as a function of the wavelength $|\bm k_1|$. 
Top panel shows the result of $\langle\theta_{\bm k_1}\theta_{\bm k_2}\theta_{\bm k_3}^{2}\rangle$ and bottom panel shows the result of $\langle e^{i(\theta_{\bm k_1}+\theta_{\bm k_2}-\theta_{\bm k_3})}\rangle$.
The three point styles distinguish the three open angels $\alpha$ .
The dashed lines show analytical prediction to the $\mathcal O(V^{-1/2})$ order and dotted lines to the $\mathcal O(V^{-1})$ order. 
The errorbars are measured from downsample method.
The grey shadow regions indicate the fluctuation of the prediction among the downsamples.
\label{fig:xiphph3pt1}}
\end{figure}
Fig. \ref{fig:xiphph3pt1} shows the 3-pt phase correlation functions at redshift $z=0$. For brevity, we only show
configurations that $|\bm k_1|$ and $|\bm k_2|$ fall within the same range ($|\bm k_1|\approx|\bm k_2|$), and the bin width $\sim 0.1 h{\rm Mpc}^{-1}$. 
We shown configurations with three open angles ($20^\circ,60^\circ,120^\circ$) between $\bm k_1$ and $\bm k_2$.  
The correlation function are measured within each $|\bm k_1|$ and
$\alpha$ bins.  We split the $k$ modes into 8 subsamples
to estimate the measurement errorbars.  Again we detect significant
non-Gaussianity, which agrees with the 
leading order theory prediction excellently. 

\subsection{Impact of phase-modulus cross-correlation}
To test the influence of disrupting the modulus-phase correlation on the polyspectrum, we make new realizations from the simulation results by randomizing the modulus or phases.
That is, for the Fourier coefficients $f(\boldsymbol{k})\equiv A_{\boldsymbol{k}} e^{i\theta_{\boldsymbol{k}}}$, to randomize phases we replace the phase $\theta_{\boldsymbol{k}}$ for each pixel with an independent random phase $\theta^\prime_{\boldsymbol{k}}$  while keeping the modulus fixed.
The first new field $f^\prime(\boldsymbol{k})=A_{\boldsymbol{k}} e^{i\theta^\prime_{\boldsymbol{k}}}$ is then Gaussian Random Field (GRF) as the modulus are independently Rayleigh distributed (Eq. \ref{eq:jpdfAk}).
To realize the randomization of the modulus, we keep the phases fixed.
And we replace $A_{\boldsymbol{k}}$ of each pixel with an independent random modulus $A^\prime_{\boldsymbol{k}}$ sampled from the Rayleigh distribution (Eq. \ref{eq:jpdfAkG}). 
The second new field $f^{\prime\prime}(\boldsymbol{k})=A^{\prime}_{\boldsymbol{k}} e^{i\theta_{\boldsymbol{k}}}$ then retain the phase information with phase-modulus correlation removed.
We then measure the bispectrum from the two new fields and compare to
the original field.

\begin{figure} 
%\hspace{-2cm}
\centering
\includegraphics[width=9.25cm]{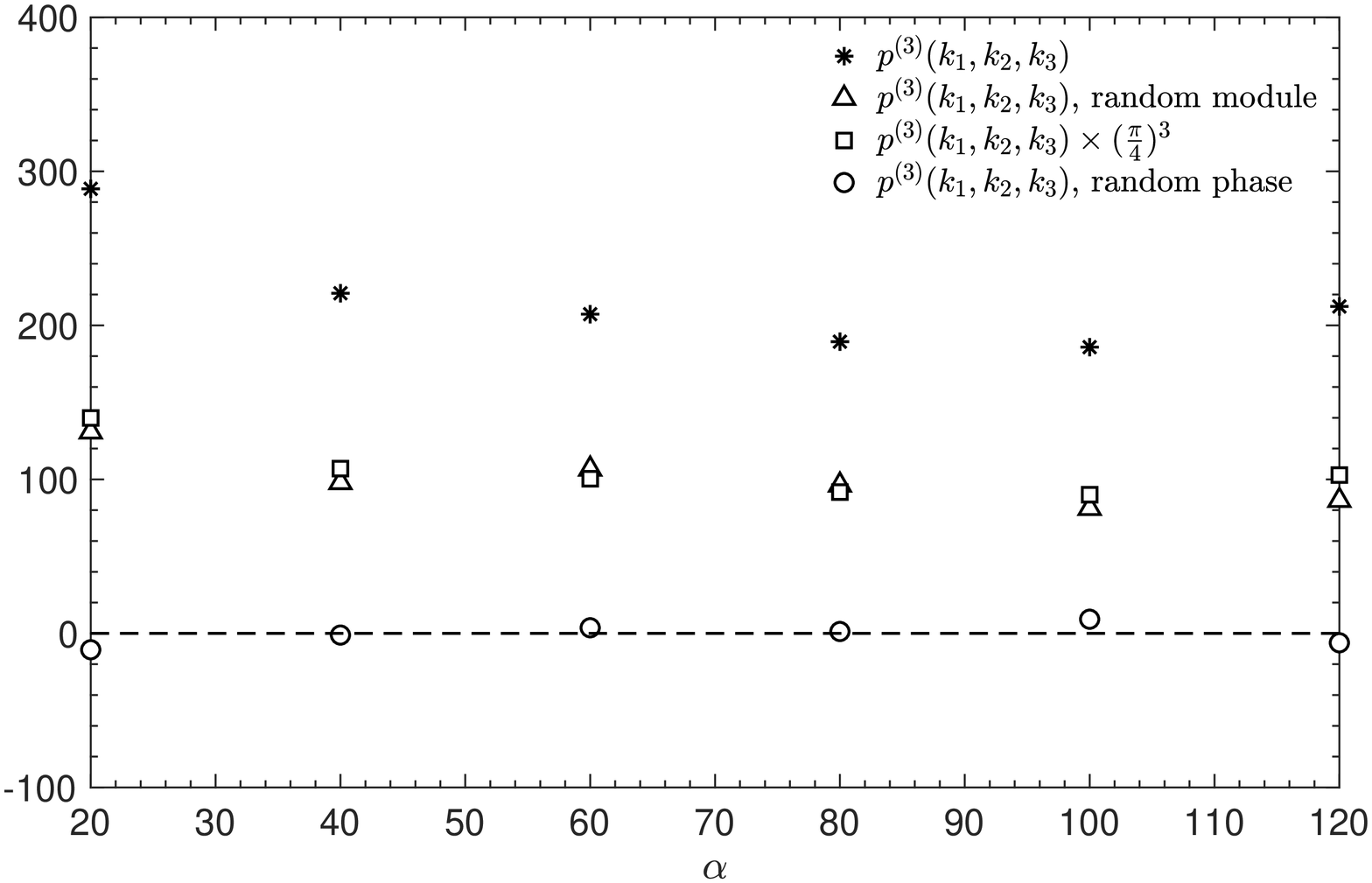}
%\vspace{-1.5cm}
\caption{The reduced bispectrum from the density field at $z=0$ (star dots), the field after randomization of the Fourier modulus (triangular dots) and the field after randomization of Fourier phases (round dots). The square dots denote the analytical prediction for the modulus randomized field. Zeros are expected for the phase randomized field (the dashed line).
\label{fig:randomized}}
\end{figure}

The analytical prediction of the influence of the bispectrum for disrupting phase-modulus correlation is equation \eqref{eq:newbs}, that is, a factor of $(\frac{\pi}{4})^3\simeq0.5$ decrease in the bispectrum.  
On the other hand, the phase randomized field is Gaussian (to first order), vanished bispectrums are then predicted for this field.
%And integration over Eq. \ref{eq:3ptphjpdf1_2nd} indicates that the second order correction is zero for both cases.
In Fig. \ref{fig:randomized} the reduced bispectrum from the density field at $z=0$ and that from the randomized fields are plotted. 
The round and triangular data points show the results of the phase randomized field and  modulus randomized field, respectively. 
The bispectrums are measured from the configuration that $|\bm k_1|$ and $|\bm k_2|$ lie in range $[0.5,0.6]\ h^{-1}{\rm Mpc}/h$ and open angle $\alpha$ (between $\bm k_1,\bm k_2$) lie in ranges with bin width $2\degr$. 
The x-axis are the average open angles within each $\alpha$ bins.
The dashed line (zero value line) and the square points represent the prediction for the randomized fields and the star points show the result of the original field.

The results agree with the predictions for both fields, indicating again the validity of the perturbation theory and the first order approximations. 
The decrease of the bispectrum of the modulus randomized field implies the contribution of the Fourier modulus to the non-Gaussianity, in way of phase-modulus coupling.

\section{Summary}
In this work, we analyzed the two-point and three-point probability
distributions of modulus and phase of Fourier modes through
N-body simulations, and  made comparison to the theory prediction. We
further measured the phase and modulus  correlation functions, and
derived the theoretical prediction and made the comparisions. We also
investigated the phase-modulus cross-correlation. We found that it contributes
$\sim 50\%$ to the measured bispectrum. 

We found that the agreement between the simulation data and
$\mathcal{O}(V^{-1/2})$ order approximation is generally good,
especially for the three point cases due to a larger sample
size. These results are consistent with the numerical investigation of
three-point phase PDF
\citep{2004ApJ...600..553H}. We also calculates the
$\mathcal{O}(V^{-1})$ order correction, and found that they are always
negligible. The analysis of \citep{2004ApJ...600..553H} used simulation volumes
$V\leq (300h^{-1}{\rm Mpc})^3$. Our simulation volume is
$V=(600h^{-1}{\rm Mpc})^3$. Most galaxy and weak lensing surveys
have larger volume, at least of the order $(1000h^{-1}{\rm
  Mpc})^3$. This means that for the related Fourier mode analysis of these
surveys, the $\mathcal{O}(V^{-1/2})$ term is the only non-Gaussian
correction that we need to include. The same methodology can be
extended to the 2D field such as the weak lensing convergence
field and the SZ effect. We expect weaker non-Gaussianity \citep[e.g.,][]{2011MNRAS.418..145J,2014JCAP...04..004M,2011PhRvD..84b3523Y,2012PhRvD..86b3515Y,2016PhRvD..94h3520Y,2020arXiv200110765C,2007ApJ...671...14Z}, and therefore the
$\mathcal{O}(V^{-1/2})$ order term  (or equivalently the
$\mathcal{O}(f_{\rm sky}^{-1/2})$ term) will be sufficient for the 2D
Fourier mode (or harmonic mode) PDF. 

This work further demonstrates that, thanks to the central limit
theorm, non-Gaussianities in Fourier space are weaker and
simplier. Therefore data analysis in Fourier space has specific
advantange, and may be worthy of further investigation. 
   
\section*{Data availability}

All data included in this study are available upon request by contact with the corresponding author. 
     
\section*{Acknowledgements}
This work was supported by the National Science Foundation of China
(No. 11621303, 11653003).
\bibliographystyle{mnras}
\bibliography{mybib}
\bsp	% typesetting comment
\label{lastpage}
\end{document}